\documentclass[]{aa}

\def\rmit#1{{\it #1}}              
\def\specchar#1{{\sc #1}}

\def\SiI{\mbox{Si\,\specchar{i}}}
\def\CaI{\mbox{Ca\,\specchar{i}}}
\def\HeI{\mbox{He\,\specchar{i}}}
\def\CaIIH{\mbox{Ca\,\specchar{ii}\,\,H}}       

\def\ie{\rmit{i.e.}}
\def\eg{\rmit{e.g.}}

\def\arcsec{\hbox{$^{\prime\prime}$}}

\usepackage{natbib}
\bibpunct{(}{)}{;}{a}{}{,} 
\usepackage{graphicx}
\usepackage{amsmath}

\usepackage{color}
\def\bc{\color{black}}

\usepackage{hyperref}
\hypersetup{
    final=true,
    pageanchor=true,
    colorlinks=true,
    breaklinks=true,
    linkcolor=blue,
    citecolor=blue,
    urlcolor=blue,
    pdfpagemode=UseNone,
    pdftitle={Magnetic and thermodynamic structure of a decaying sunspot},
    pdfauthor={T. Felipe et al.},
    pdfsubject={Solar Physics},
    pdfkeywords={methods: data analysis - Sun: photosphere - Sun: magnetic fields - sunspots}}

\begin{document}

\title{Three-dimensional structure of a sunspot light bridge}

\author{T. Felipe\inst{\ref{inst1},\ref{inst2}}
\and M. Collados\inst{\ref{inst1},\ref{inst2}}
\and E. Khomenko\inst{\ref{inst1},\ref{inst2}}
\and C. Kuckein\inst{\ref{inst3}}
\and A. Asensio Ramos\inst{\ref{inst1},\ref{inst2}}
\and H. Balthasar\inst{\ref{inst3}}
\and T. Berkefeld\inst{\ref{inst4}}
\and C. Denker\inst{\ref{inst3}}
\and A. Feller\inst{\ref{inst5}}
\and M. Franz\inst{\ref{inst4}}
\and A. Hofmann\inst{\ref{inst3}}
\and C. Kiess\inst{\ref{inst4}}
\and A. Lagg\inst{\ref{inst5}}
\and H. Nicklas\inst{\ref{inst6}}
\and D. Orozco Su\'arez\inst{\ref{inst10}}
\and A. Pastor Yabar\inst{\ref{inst1},\ref{inst2}}
\and R. Rezaei\inst{\ref{inst1},\ref{inst2}}
\and R. Schlichenmaier\inst{\ref{inst4}}
\and D. Schmidt\inst{\ref{inst9}}
\and W. Schmidt\inst{\ref{inst4}}
\and M. Sigwarth\inst{\ref{inst4}}
\and M. Sobotka\inst{\ref{inst7}}
\and S. K. Solanki\inst{\ref{inst5},\ref{inst8}}
\and D. Soltau\inst{\ref{inst4}}
\and J. Staude\inst{\ref{inst3}}
\and K. G. Strassmeier\inst{\ref{inst3}}
\and R. Volkmer\inst{\ref{inst4}}
\and O. von der L\"uhe \inst{\ref{inst4}}
\and T. Waldmann \inst{\ref{inst4}}
}


\institute{Instituto de Astrof\'{\i}sica de Canarias, 38205, C/ V\'{\i}a L{\'a}ctea, s/n, La Laguna, Tenerife, Spain\label{inst1}
\and 
Departamento de Astrof\'{\i}sica, Universidad de La Laguna, 38205, La Laguna, Tenerife, Spain\label{inst2} 
\and 
Leibniz-Institut f\"ur Astrophysik Potsdam (AIP), An der Sternwarte 16, 14482 Potsdam, Germany \label{inst3}
\and
Kiepenheuer-Institut f\"ur Sonnenphysik, Sch\"oneckstr. 6, 79104 Freiburg, Germany\label{inst4}
\and
Max-Planck-Institut f\"ur Sonnensystemforschung, Justus-von-Liebig-Weg 3, 37077, G\"ottingen, Germany\label{inst5}
\and
Institut f\"ur Astrophysik, Georg-August-Universit\"at G\"ottingen, Friedrich-Hund-Platz 1, 37077 G\"ottingen, Germany\label{inst6}
\and
Instituto de Astrof\'isica de Andaluc\'ia (CSIC), Apdo. 3040, E-18080 Granada, Spain\label{inst10}
\and
National Solar Observatory, 3010 Coronal Loop Sunspot, NM 88349, USA\label{inst9}
\and
Astronomical Institute of the Academy of Sciences, Fri\v{c}cova 298, 25165 Ond\v{r}ejov, Czech Republic\label{inst7}
\and
School of Space Research, Kyung Hee University, Yongin, Gyeonggi Do, 446-701, Republic of Korea\label{inst8}
}

\abstract
{Active regions are the most prominent manifestations of solar magnetic fields; their generation and dissipation are fundamental problems in solar physics. {\bc Light bridges are commonly present during sunspot decay, but a comprehensive picture of their role in the removal of photospheric magnetic field is still missing.}} 
{We study the three dimensional configuration of a sunspot and in particular its light bridge during one of the last stages of its decay.}
{We present the magnetic and thermodynamical stratification inferred from full Stokes inversions of the photospheric \SiI\ 10827 \AA\ and \CaI\ 10839 \AA\ lines obtained with the GREGOR Infrared Spectrograph of the GREGOR telescope at Observatorio del Teide, Tenerife, Spain. The analysis is complemented by a study of continuum images covering the disk passage of the active region, which are provided by the Helioseismic and Magnetic Imager on board the Solar Dynamics Observatory.}
{The sunspot shows a light bridge with penumbral continuum intensity that separates the central umbra from a smaller umbra. We find that in this region the magnetic field lines form a canopy with lower magnetic field strength in the inner part. The photospheric light bridge is dominated by gas pressure (high-$\beta$), as opposed to the surrounding umbra where the magnetic pressure is higher. A convective flow is observed in the light bridge. This flow is able to bend the magnetic field lines and to produce field reversals. The field lines close above the light bridge and become as vertical and strong as in the surrounding umbra. We conclude that it develops because of two highly magnetized regions which come closer during the sunspot evolution.}
{}

\keywords{Methods: observational -- methods: data analysis -- Sun: photosphere -- Sun: magnetic fields -- sunspots -- Sun: activity}

\maketitle


\section{Introduction}

The determination of the magnetic topology of sunspots has been an important topic in solar physics during the last decades. A global scale picture has emerged, showing that sunspots are formed by a flux tube with a strong and vertical magnetic field at the axis, in the umbral center, which becomes weaker and more horizontal at radial distances farther from the flux tube's axis \citep[\eg,][]{Tiwari+etal2015}. The improvement in the resolution of the observations has revealed a more complex picture, where fine structure is present in the umbra and penumbra of sunspots \citep[see][for a review]{Solanki2003, Borrero+Ichimoto2011}. The magnetic structure is a result of the close interplay between plasma motions and magnetic fields, and it is subject to significant changes during the evolution of the sunspot.

\begin{figure*}[!ht] 
 \centering
 \includegraphics[width=14cm]{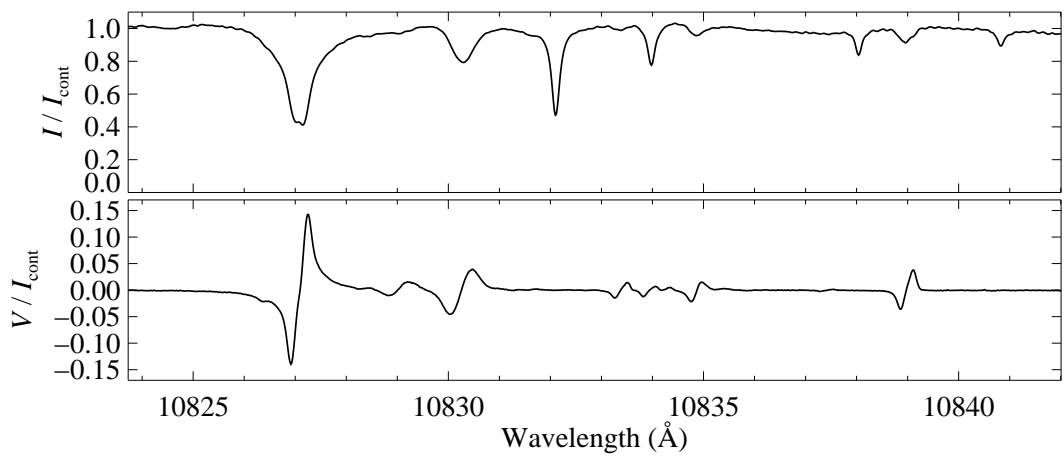}
  \caption{Stokes I (top panel) and Stokes V (bottom panel) at position C on Fig. \ref{fig:mapa_GREGOR} normalized to quiet Sun continuum intensity.}
  \label{fig:spectra}
\end{figure*}

The decay of a sunspot is a slow process, which usually takes place over several days. The mechanisms leading to the disappearance of the sunspot's magnetic flux are still not fully understood and remain one of the most challenging problems of sunspot physics. Most sunspots exhibit a large-scale outflow called ``moat flow'' \citep{Meyer+etal1974} beyond their visible outline, whose physical origin differs from that of the Evershed flow \citep{Lohner-Bottcher+Schlichenmaier2013}. It was first detected spectroscopically \citep{Sheeley1972} and then by tracking Moving Magnetic Features \citep[MMFs,][]{Harvey+Harvey1973}. MMFs with the same polarity as the umbra have been associated with sunspot decay \citep[\eg,][]{MartinezPillet2002} because they extract net flux from the spot \citep{Zhang+etal1992}. They disperse this flux, which is then removed from the photosphere. However, the flux removal mechanism remains elusive. MMFs move through the moat flow, accumulating at the edge of the moat region. On the other hand, pairs of MMFs with opposite polarities do not contribute to the flux removal since they do not carry net flux. \citet{Deng+etal2007} studied the flow field in a decaying sunspot and concluded that the mechanisms contributing to the decay are the fragmentation of the umbra, cancellation of MMFs that encounter opposite polarity areas at the moat's outer boundary, and flux transport by MMFs to regions with the same polarity as the sunspot. After the disappearance of the sunspot, the flow pattern in that location exhibits significant differences with flows at other quiet Sun regions \citep{Verma+etal2012}. \citet{BellotRubio+etal2008} studied the decay of a penumbra using spectropolarimetric data. Blue-shifted finger-like structures of weak and horizontal magnetic field extending outward from the border of the umbra were identified. They proposed that these structures are penumbral field lines that rise to the chromosphere and contribute to the disappearance of the penumbra at photospheric levels.

Sunspot light bridges can be formed due to the fragmentation of the umbra in the decay phase or during the merging of different magnetized areas during the formation of the sunspot in complex active regions \citep{Bray+Loughhead1964, GarciadelaRosa1987}. At the last stages of a sunspot, as a consequence of the splitting of the umbra the photospheric-like conditions are recovered, and a granulation pattern similar to the quiet Sun is found in light bridges \citep{Vazquez1973}, although the light bridge convection cells do differ significantly from normal granules \citep{Lagg+etal2014}. Understanding the decay of sunspots requires the study of the magnetic and dynamical structure of light bridges. Previous studies have classified light bridges based on their morphological properties. Light bridges separating two umbral regions are called ``strong light bridges'' \citep[\eg,][]{Sobotka+etal1993, Sobotka+etal1994,Jurcak+etal2006, Rimmele2008}. Their brightness is similar to that of the penumbra and they usually appear between two regions with the same polarity. ``Faint light bridges'' \citep[\eg][]{Lites+etal1991, Sobotka+etal1993} are elongated bright structures which penetrate the umbra. They are composed of rows of bright grains with sizes comparable to umbral dots. Light bridges can exhibit different internal structure. Most of them show several segments along their length that resemble granules \citep{Berger+Berdyugina2003}, while some light bridges are similar to the bright filaments seen in the penumbra \citep{Lites+etal2004}. All types show a weakened magnetic field strength relative to the surrounding umbra and more inclined field lines \citep[\eg][]{Beckers+Schroter1969, Lites+etal1991, Leka1997}, which in some case even exhibit a polarity reversal \citep{Lagg+etal2014}. Observations are consistent with a reduced field strength in the photosphere with a magnetic canopy extending from both sides of the light bridge and merging at the top \citep{Jurcak+etal2006, Lagg+etal2014}. The velocity field measured in light bridges shows evidence of their convective origin \citep{Rimmele1997, Hirzberger+etal2002, Lagg+etal2014}. Recently, \citet{Schlichenmaier+etal2016} have also found evidence that the umbral magnetic field wraps around light bridges. They defined a new type, the plateau light bridge, with `Y' shaped dark canals that resemble penumbral grains and suggest the presence of inclined magnetic fields. Several dynamic phenomena have been detected at the chromosphere above light bridges, including recurrent plasma ejections visible in H$\alpha$ \citep{Roy1973,Asai+etal2001}, \CaIIH\ \citep{Shimizu+etal2009} and other bands \citep[AIA 1600 and 1700 \AA, IRIS 1330 and 1400 \AA,][]{Toriumi+etal2015}, brightness enhancements in TRACE 1600 \AA\ band \citep{Berger+Berdyugina2003}, or fan-shaped jets observed in H$\alpha$ \citep{Robustini+etal2016}. All these processes are considered to be caused by the interaction of the light bridge magnetic field with the surrounding atmosphere through magnetic reconnection events.

Sunspot evolution is driven by the interaction of magnetic fields and plasma. The relevance of the plasma motions in the molding of the magnetic field is quantified by the plasma-$\beta$ parameter \citep[\eg,][]{Gary2001}. It is defined as the ratio between the gas pressure, $P_\mathrm{g}$, and the magnetic pressure, $P_\mathrm{m}$, \ie, $\beta=P_\mathrm{g}/P_\mathrm{m}=2\mu_0P_\mathrm{g}/B^2$, where $\mu_0$ is the magnetic permeability and $B$ is the magnetic field strength. In those regions where $\beta\gg 1$ the plasma dominates over the magnetic field, which can be bent and dragged following plasma motions. On the other hand, where $\beta\ll 1$ the magnetic field is not influenced by the plasma, and motions can only occur along field lines. In these regions the magnetic field acquires a force-free configuration \citep[see Sect. 2.4 from][]{Borrero+Ichimoto2011}. In sunspot atmospheres, the umbra shows small $\beta$ values ($\beta <1$) at the low photosphere and higher values ($\beta >1$) as we move towards the edges of the spot \citep[\ie,][]{Mathew+etal2004}, although for at least one other sunspot, plasma  $\beta>1$ everywhere in the sunspot's lower atmosphere \citep{Solanki+etal1993}. At higher layers the atmosphere is dominated by the magnetic field ($\beta\ll 1$).

In this study, we aim at analyzing the thermodynamic and magnetic structure of a sunspot during its decay, a few days before it completely disappeared. We concentrate on the deep photosphere, where the magnetic field configuration is more dependent on the plasma dynamics. In Sect. \ref{sect:observations}, we present the data set and qualitatively describe the evolution of the active region during its disk passage. Section \ref{sect:inversion} shows the inversion techniques used in the analysis, and the results are described in Sect. \ref{sect:results}. Finally, conclusions are outlined in Sect. \ref{sect:conclusions}.

\section{Observations}
\label{sect:observations}
Spectropolarimetric data of the active region NOAA 12096 were obtained on 2014 June 27 from 8:28 to 9:13 UT with the 1.5-meter GREGOR solar telescope \citep{Schmidt+etal2012} at the Observatorio del Teide, Tenerife, Spain. The center of the analyzed region was located at the solar position $x=-184''$, $y=100''$, which corresponds to heliocentric angle of $\Theta =12.4^o$ ($\mu=\cos\Theta=0.98$).

The full data set consists on 10 maps of the four Stokes parameters, each of them covering a scanned area of $61''\times 7.8''$, acquired with the GREGOR Infrared Spectrograph \citep[GRIS,][]{Collados+etal2012}. The pixel size along the slit was 0.135\arcsec. For the construction of the maps, 60 consecutive slit positions were included by shifting the slit in increments of 0.135\arcsec. The number of accumulations per recording was 10 with an integration time of 40 ms each. A full map was completed in 4.5 min. The wavelength scale was calibrated on an absolute scale \citep{MartinezPillet+etal1997}, correcting for several systematic effects such as Earth's orbital motions and the solar gravitational redshift \citep[see Appendices A and B of][]{Kuckein+etal2012b}. The spectral range spans from 10823.73 to 10842.01 \AA\ with a spectral sampling of 18.1 m\AA\ pixel$^{-1}$. Figure \ref{fig:spectra} shows the Stokes $I$ and $V$ parameters at location C near the umbra/penumbra boundary (see Fig. \ref{fig:mapa_GREGOR}). The spectral range of the spectropolarimeter comprises several spectral lines of interest, such as the photospheric \SiI\ 10827 \AA\ and \CaI\ 10839 \AA\ lines and the chromospheric \HeI\ 10830 \AA\ triplet. In this study we focus on the analysis of the two aforementioned photospheric lines.

Standard procedures for dark and flat-field correction and polarimetric calibration \citep{Collados1999, Collados2003} were applied to all images. The telescope calibration unit \citep{Hofmann+etal2012} was used for the polarimetric calibration. The GREGOR Adaptative Optics System \citep[GAOS,][]{Berkefeld+etal2012} was locked on nearby structures, improving substantially the image quality. GREGOR's altitude-azimuthal mount introduced an image rotation \citep{Volkmer+etal2012}. Obtaining co-spatial maps required the application of some correction to compensate this rotation, which amounts to $11^\circ$ during the whole period of the observations. However, since the maps do not show significant evolution during the observed 45 min, we have decided to focus the analysis on each individual map so that a temporal series of co-spatial maps is not necessary. The image rotation during the acquisition of each map (slightly higher than $1^\circ$) has been neglected.  

\begin{figure*}[!ht] 
 \centering
 \includegraphics[width=16cm]{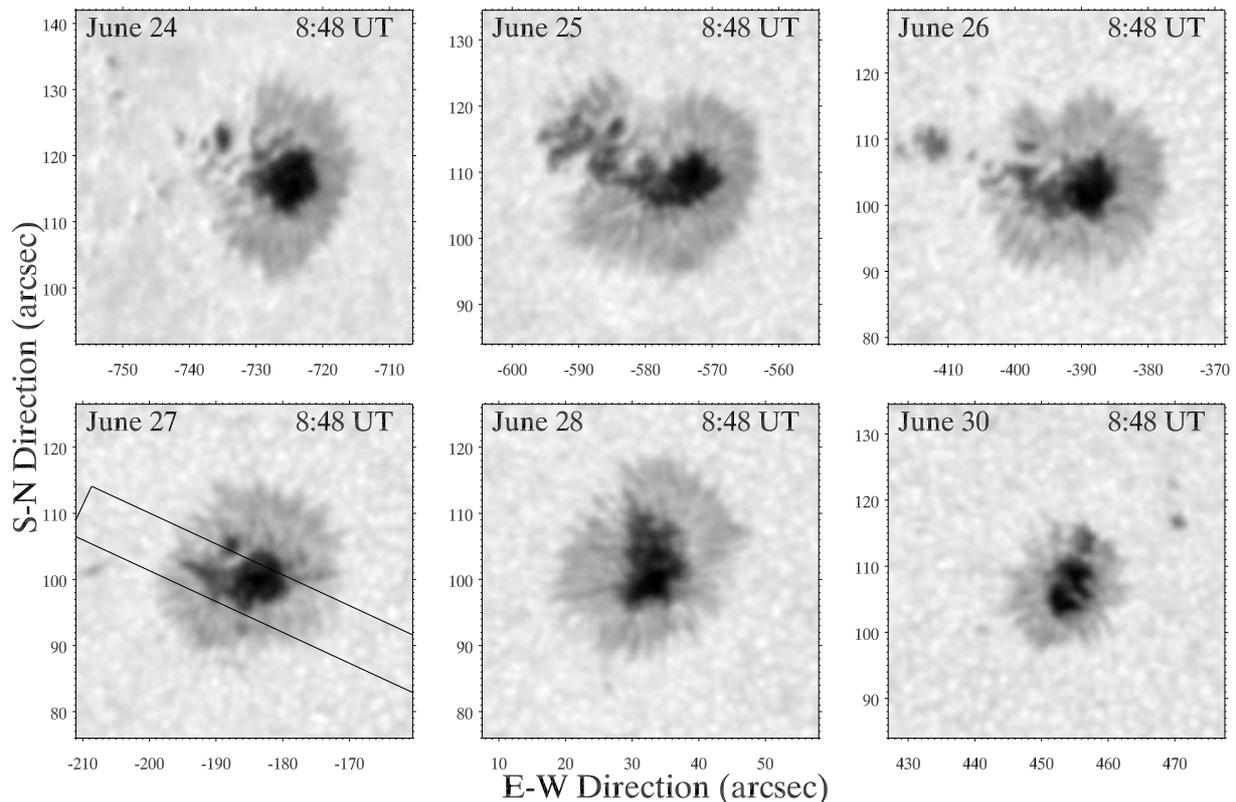}
  \caption{HMI continuum images of the evolution of the active region NOAA 12096 between 2014 June 24\,--\,30. Date and time of the images are indicated at the top of each panel. The tilted rectangle in the bottom left panel indicates the FOV covered by the GRIS observations.}
  \label{fig:evolution_HMI}
\end{figure*}

\subsection{PSF and stray light corrections}
\label{sect:PSF}

The spatial resolution of solar ground-based observations is limited by the turbulence in the Earth atmosphere and by the optical properties of the telescope \citep[\eg,][]{MartinezPillet1992, Chae+etal1998, WedemeyerBohm2008}. The distortion that these effects produce on the observations is characterized by the Point Spread Function (PSF), whose contributions are commonly approximated by analytical functions. The PSFs are usually dominated by an Airy pattern and, as a consequence of its tails, light observed at the focal plane is contaminated by light coming from distant spatial locations. This spurious light is known as "stray light". Deconvolution techniques are a common approach for correcting this effect. In the case of two dimensional observations (broad-band imaging or narrow-band spectroscopy) an effective instantaneous PSF can be estimated observationally \citep[\eg,][]{vonderLuhe1993, vanNoort+etal2005}. These techniques are not applicable to maps obtained from the scanning of adjacent spatial steps using a slit-spectrograph, since at each scan the PSF includes contributions from surrounding regions not covered by the one-dimensional slit. Instead, the deconvolution scheme is applied to each monochromatic map using a static PSF \citep[\eg][]{Beck+etal2011,QuinteroNoda+etal2015}. The correction for stray light using deconvolution techniques has produced observations with unprecedented spatial resolution and has allowed the study of the Sun at finer scales \citep[\eg,][]{Scharmer+etal2011, Joshi+etal2011, Henriques2012}.

\citet{Schlichenmaier+Franz2013} have claimed that the correction for scattered light can produce artifacts on the velocity measurements if the PSF of the stray light is not well characterized. In this work, we have empirically determined the PSF of the GRIS instrument from observations of the Mercury transit in front of the Sun on 2016 May 9. During the observations, weather conditions were not optimum, with some clouds at the level of the observatory. A wavelength-integrated map of Mercury transiting the solar disk was obtained in the 1.56 $\mu$m region using GRIS polarimetric mode. The scan was composed of 300 consecutive recordings with a 0.135\arcsec\ step size. The integration time per individual image was 100 ms and the number of accumulations was 3. 

A synthetic circular image simulating the undegraded Mercury covering a homogeneous solar surface was created and convolved with a PSF profile given by

\begin{equation}
\phi (r)=\alpha G_1(r,\sigma_1)+(1-\alpha)G_2(r,\sigma_2),
\label{eq:PSF}
\end{equation}

\noindent where $G_1$ and $G_2$ are two Gaussians with the rms widths given by $\sigma_1$ and $\sigma_2$, respectively. The first Gaussian represents the spatial resolution of the telescope (including the seeing) and the second Gaussian accounts for the instrumental stray light. The parameter $\alpha$ is the relative contribution of each Gaussian. The parameters $\alpha$, $\sigma_1$, and $\sigma_2$ were obtained from a non-linear fit of the synthetic convolved circular image to the observed Mercury map. 

The fit yields $\alpha=0.584\pm0.001$, $\sigma_1=1.385\pm0.001$\arcsec, and $\sigma_2=5.030\pm0.001$\arcsec. The FWHM of the first Gaussian (3.26\arcsec ) is consistent with the angular resolution of the observations, which were seeing-limited to 3.3\arcsec. This fact supports the assumption of the PSF proposed in Eq. \ref{eq:PSF} and allow us to reasonably argue that the second Gaussian gives the stray light contamination, with a FWHM of 11.8\arcsec\ and a relative weight of approximately 42\%. 

The spectropolarimetric maps of the sunspot were deconvolved using a principal component analysis regularization \citep{RuizCobo+AsensioRamos2013}. We used a PSF given by Eq. \ref{eq:PSF}, but substituted $G_1(r,\sigma_1)$ by the Airy profile of the GREGOR telescope with the radius of the aperture corresponding to the Fried parameter $r_0$. During the scanning of the map analysed in this paper, the average value for the whole scan at 10830 \AA\ was $r_0=50.7$ cm. Due to the unfortunate seeing conditions during the Mercury transit, the small-scales properties of the stray light can be masked under the seeing contribution to the $G_1$ component of the PSF.

\subsection{Evolution of the sunspot}
\label{sect:evolution}

Continuum images obtained with the Helioseismic and Magnetic Imager \citep[HMI,][]{Schou+etal2012} on board the Solar Dynamic Observatory (SDO) have been examined in order to discuss the temporal evolution over a few days of the sunspot NOAA 12096 during its decay. HMI provides full disk observations with 4096$\times$4096 pixels and a pixel size of 0.5\arcsec. A selection of one image every 12 minutes between 2014 June 23 UT 16:00 and 2014 July 1 UT 23:48 was extracted from the HMI database. The full temporal series includes 1000 time steps. 

Figure \ref{fig:evolution_HMI} shows a sample of six continuum images with a cadence of one day, except for the last panel, which corresponds to two days after the previous snapshot. A movie with the evolution of the sunspot during the time span indicated in the previous paragraph can be found in the online material. Between June 23 and June 26 the sunspot group was classified with a Hale class $\alpha$. During this time, part of the umbra separates from the center of the sunspot and moves towards the east. On June 26 at 08:48 UT, this region is completely detached from the outer penumbra of the sunspot. At the same time, a darker penumbral region (which appears as a dark pore-like structure at $x=-735$, $y=122.5$ on June 24 at 08:48 UT) moves towards the umbra. On June 27 this region is separated from the umbral core and from a ``spike-like'' extension of the main umbra by a zone with penumbral brightness. It forms the ``strong light bridge'' that is analysed in this study. Our GRIS observations were taken at that moment. The area covered by them is delineated by solid black lines in the bottom left panel of Fig. \ref{fig:evolution_HMI}. At this time, the group was classified as a $\beta$-region, although the region under investigation only shows negative polarity. The dark pore-like feature then merges with the rest of the umbra, forming an extended umbra with internal structure. At the last stages of the evolution of the sunspot, the umbra is divided by a light-bridge into two separated dark regions with the same polarity (see last panel of Fig. \ref{fig:evolution_HMI}), and finally it loses the surrounding penumbra. A C-class flare on June 24 at 20:56 UT was associated with this active region.

\begin{figure*}[!ht] 
 \centering
 \includegraphics[width=14cm]{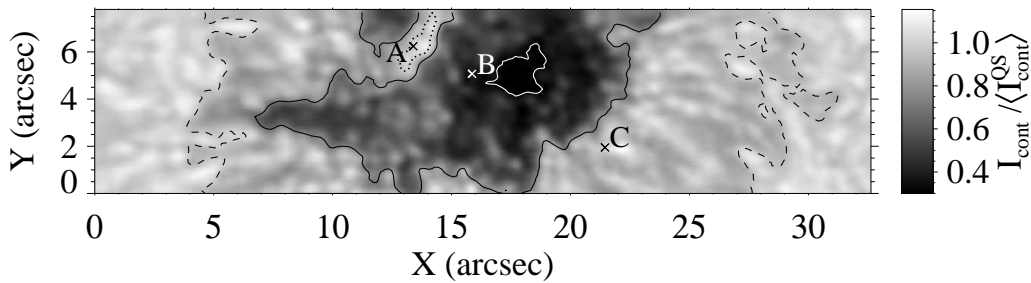}
  \caption{Continuum intensity map normalized to the average continuum intensity in the quiet Sun $\langle I^\mathrm{QS}_\mathrm{cont}\rangle$ obtained with GRIS on June 27 between 08:46 and 08:50 UT. Points $A$, $B$, and $C$ correspond to the locations used for Figs. \ref{fig:spectra}, \ref{fig:RFs}, \ref{fig:inversion_ff}, and \ref{fig:inversion_umbra}. The contour lines mark the boundaries of the masks used for the analysis in Sect. \ref{sect:results}: ``Dark umbra'' inside the white line, ``faint umbra'' delimited by the black solid line and the white line, ``penumbra'' between the black solid line and the dashed lines, and the ``light bridge'' inside the dotted line.}
  \label{fig:mapa_GREGOR}
\end{figure*}

A section of the field-of-view (FOV) covered by the GRIS observations is shown in Fig. \ref{fig:mapa_GREGOR}. It corresponds to the continuum intensity at a wavelength close to the \SiI\ 10827 \AA\ line. This is the fifth map of the series acquired with GRIS and it was obtained between 08:46 and 08:50 UT. Since the ten maps of the series provide similar results, from this point onwards all figures and discussion refer to this time step. The contours overplotted on the map delimit the boundaries of some masks, which have been applied for the analysis of the data. The region inside the white line includes the darkest part of the umbra and in the following this mask will be referred as ``dark umbra''. What we term the ``faint umbra'' corresponds to the area between the white line and the black solid line and also the small umbra around $x=13''$ and $y=7''$. The ``penumbra'' is delimited by the black dashed and solid lines. Finally, the area surrounded by the dotted line between the dark pore-like feature and the umbra will be called ``light bridge''.

\section{Spectral lines inversion and data analysis}
\label{sect:inversion}

The photospheric \SiI\ 10827 \AA\ and \CaI\ 10839 \AA\ lines were inverted simultaneously using the ``Stokes Inversion based on Response functions'' (SIR) code \citep{RuizCobo+delToroIniesta1992}. SIR yields the thermal, dynamic, and magnetic depth-dependent stratification of the atmosphere that best matches the observed Stokes profiles. The solution is achieved after an iterative process, where the numerical integration of the radiative transfer equation is performed under the assumption of local thermodynamic equilibrium and using the spectral line response functions (RFs), which represent the partial derivatives of the Stokes parameters with respect to the various atmospheric parameters (at different depths).

\begin{figure}[!ht] 
 \centering
 \includegraphics[width=9cm]{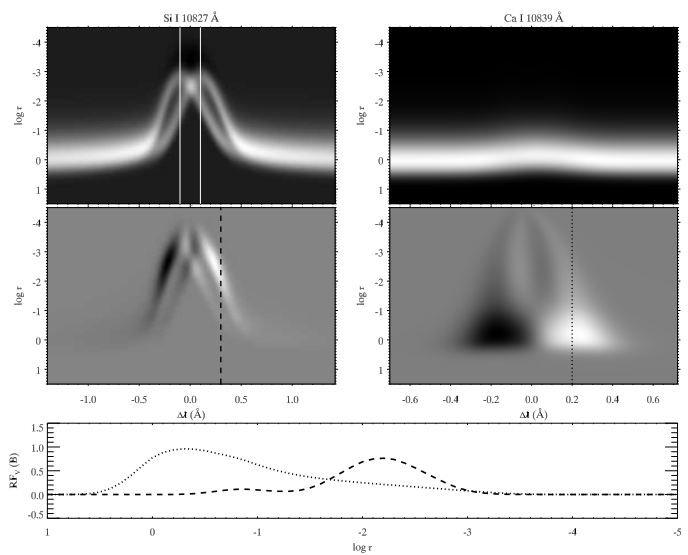}
  \caption{Normalized response functions of Stokes I to temperature (top panels) and Stokes V to the magnetic field strength (middle and bottom panels) for the atmosphere at location $A$. Top panels: response functions of Stokes I to temperature of the \SiI\ 10827 \AA\ line (left panel) and \CaI\ 10839 \AA\ line (right panel). The region between the vertical while solid lines has been neglected for the inversions. Middle panels: response functions of Stokes V to magnetic field of the \SiI\ 10827 \AA\ line (left panel) and \CaI\ 10839 \AA\ line (right panel). Bottom panels: variation of the response functions of Stokes V to magnetic field of the \SiI\ 10827 \AA\ line (dashed line) and the \CaI\ 10839 \AA\ line (dotted line) with $\log\tau$ at the wavelengths indicated by the dashed and dotted lines in the middle panels, respectively.}
  \label{fig:RFs}
\end{figure}

The \SiI\ 10827 \AA\ line has been used for probing the photosphere in many solar studies \citep[\eg,][]{Centeno+etal2006, Bloomfield+etal2007, Felipe+etal2010b, Kuckein+etal2012a}. Its proximity to the chromospheric \HeI\ 10830 \AA\ triplet makes this spectral region an ideal testing ground for simultaneous analysis of the photosphere and chromosphere. On the other hand, the \CaI\ 10839 \AA\ line, located a few {\AA}ngstr{\"o}m towards the red, has been widely overlooked \citep[though see][]{Joshi+etal2016, Balthasar+etal2016}. The broad spectral range of GRIS allows the simultaneous observation of all these spectral lines. The inclusion of the \CaI\ 10839 \AA\ line in the analysis rewards us with a better sampling of the deeper atmospheric layers, as can be seen in its response functions. Figure \ref{fig:RFs} shows the response functions of Stokes $I$ to the temperature and Stokes $V$ to the magnetic field strength for the \SiI\ 10827 \AA\ and \CaI\ 10839 \AA\ lines at location $A$ in Fig. \ref{fig:mapa_GREGOR}. The top panels show that the \SiI\ line is senstitive to heights up to $\log\tau=-3$, while the highest sensitivity of core of the \CaI\ line is slightly above the height of the continuum formation. The Stokes $V$ signal of the \SiI\ 10827 \AA\ line is sensitive to the magnetic field in a spectral range of 1 \AA\ centered at the core of the line. The maximum value of the RF at a specific wavelength is found at $\log\tau=-2.7$, where $\tau$ corresponds to the continuum optical depth at 5000~\AA. On the other hand, the peak of the \CaI\ 10839 \AA\ RF is located at a larger optical depth, around $\log\tau=-0.4$. The aforementioned results refer to the RFs at location A. We have computed the RFs to magnetic field perturbations at several locations, including quiet Sun, penumbral, and umbral regions in order to retrieve an estimate of the average formation height of the \CaI\ and \SiI\ lines. The analysis shows that their highest sensitivity to the magnetic field is found around $\log\tau=-0.5$ and $-2.2$, respectively. We estimate that the RFs of the \CaI\ 10839 \AA\ line are sufficiently high to retrieve reliable inversions of the data for $\log\tau$ lower than +0.3. The upper limit, where the atmospheric parameters obtained from the inversions are trustworthy, is given by the RFs of the \SiI\ 10827 \AA\ line. Thus, in the following, all the results derived from the simultaneous inversions of both lines will be restricted to the $\log\tau$ range between $+0.3$ and $-4.0$.

The SIR code requires an initial guess for the atmospheric model. We choose the cool umbral model from \citet{Collados+etal1994}, covering the optical depth range $+1.4\ge \log\tau\ge -5.0$. The temperature stratification $T(\tau)$ was inverted using five nodes, three nodes were used for magnetic field strength $B(\tau)$ and magnetic field inclination $\gamma(\tau)$, and two nodes for line-of-sight (LOS) velocity $v(\tau)$. The magnetic field azimuth $\chi$ was set to be independent with height, so that a constant azimuth was imposed in the initial model and only one node was selected for it.

The solar abundances were taken from \citet{Asplund+etal2009}, while the atomic data for the \SiI\ 10827 \AA\ line were obtained from \citet{Borrero+etal2003}. Since the core of the \SiI\ 10827 \AA\ line is formed in non-LTE conditions, a spectral region with a 0.2 \AA\ width centered at the peak of the line has been neglected for the inversions (see top left panel of Fig. \ref{fig:RFs}). This mask has only been applied to the intensity profile and the spectral region neglected has been chosen based on a comparison between a quiet Sun reference atmosphere \citep[HSRA,][]{Gingerich+etal1971} and the temperature stratification retrieved from inversions of the \SiI\ 10827 \AA\ line from the FTS atlas \citep{Kurucz+etal1984} with different masks. We have assumed that Stokes Q, U, and V are not affected by non-LTE conditions. The atomic parameters of the \CaI\ 10839 \AA\ line were determined semi-empirically. The line was synthesized iteratively in the HSRA atmosphere changing the values of the enhancement factor to the van der Waals coefficient $E$ and the logarithm of the multiplicity of the level times the oscillator strength $\log(gf)$. This procedure was repeated until an agreement with the observed intensity in the \CaI\ 10839 \AA\ line from the FTS atlas was achieved. For the abundances used in this study, we find $E=1.0$ and $\log(gf)=0.10$. The different values of the macroturbulence introduced in the HSRA model used for the synthesis can lead to the determination of different $E$ and $\log(gf)$ parameters. We set the macroturbulence to 1.7 km s$^{-1}$, since with this value and the atomic parameters of the \SiI\ 10827 \AA\ line from \citet{Borrero+etal2003}, the intensity profile of the \SiI\ line synthesized in the HSRA model reproduces that from the FTS atlas.

\begin{figure*}[!ht] 
 \centering
 \includegraphics[width=16cm]{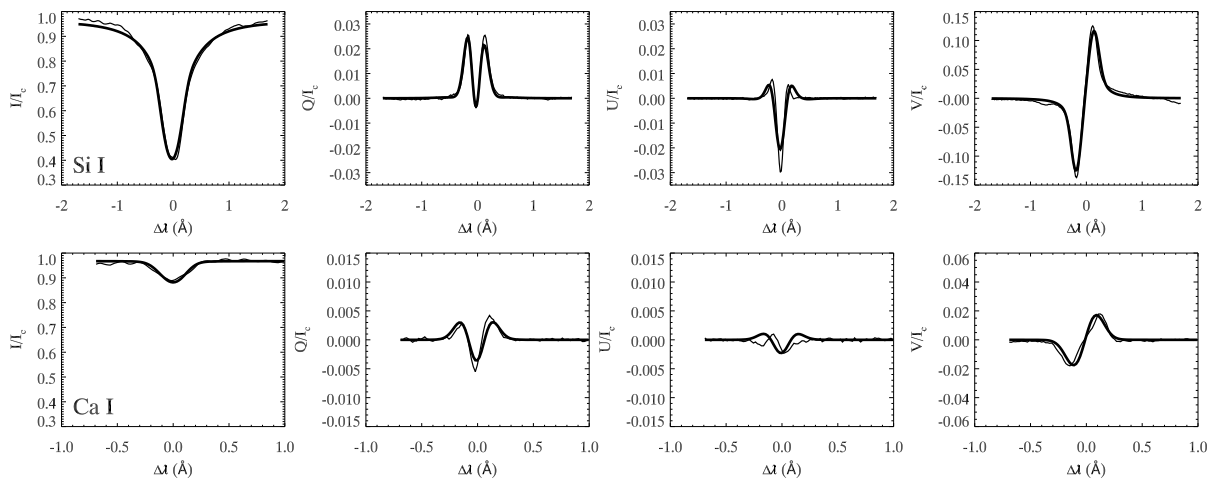}
  \caption{Stokes parameters at location $A$ (light bridge) in Fig. \ref{fig:mapa_GREGOR} for the \SiI\ 10827 \AA\ line (top panels) and the \CaI\ 10839 \AA\ line (bottom panel). Thin lines correspond to the observed profiles and thick lines to the fit obtained with SIR. The wavelength $\Delta\lambda=0$ refers to the line core. The physical parameters inferred from the inversion at $\log\tau=-0.5$ are $v=-0.01$ km s$^{-1}$, $B=568$ G, $\gamma=107^{\circ}$, and $\chi=240^{\circ}$, and at $\log\tau=-2.2$ are $v=-0.62$ km s$^{-1}$, $B=1339$ G, $\gamma=151^{\circ}$, and $\chi=190^{\circ}$.}
  \label{fig:inversion_ff}
\end{figure*}

\begin{figure*}[!ht] 
 \centering
 \includegraphics[width=16cm]{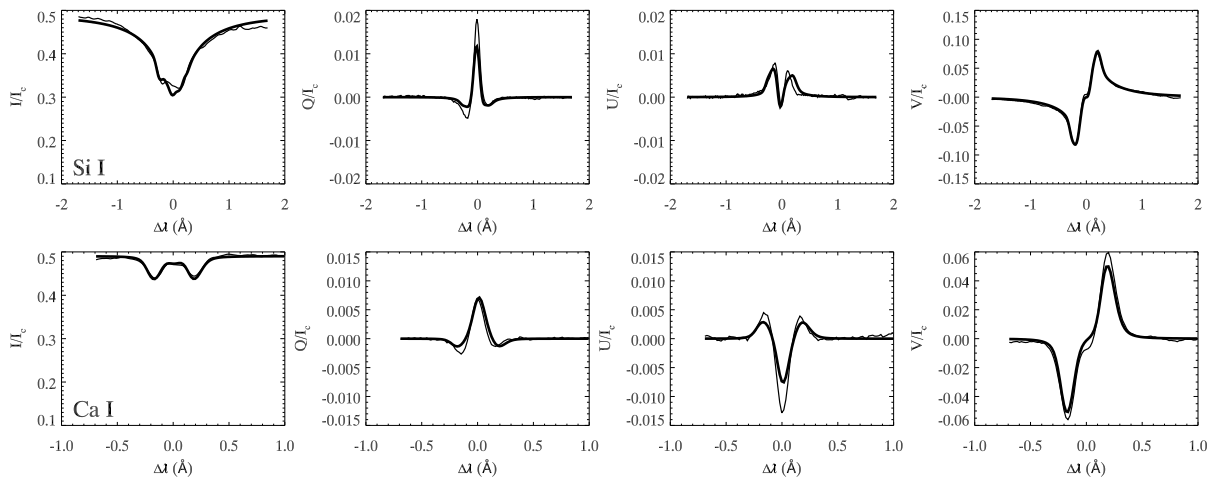}
  \caption{Same as Fig. \ref{fig:inversion_ff} but for location $B$ (umbra) in Fig. \ref{fig:mapa_GREGOR}, which corresponds to the faint umbra of the sunspot. The physical parameters inferred from the inversion at $\log\tau=-0.5$ are $v=0.43$ km s$^{-1}$, $B=2203$ G, $\gamma=160^{\circ}$, and $\chi=240^{\circ}$, and at $\log\tau=-2.2$ are $v=-0.29$ km s$^{-1}$, $B=2224$ G, $\gamma=158^{\circ}$, and $\chi=240^{\circ}$. }
  \label{fig:inversion_umbra}
\end{figure*}

Figures \ref{fig:inversion_ff} and \ref{fig:inversion_umbra} show a comparison between the observed Stokes profiles (thin line) and the results from the simultaneous inversion of the two photospheric lines (thick line) at two different positions. Each figure is composed of eight panels which represent the Stokes $I$, $Q$, $U$, and $V$ parameters for the \SiI\ 10827 \AA\ line (top panels) and the \CaI\ 10839 \AA\ line (bottom panels). Figure \ref{fig:inversion_ff} shows the profiles at a point in the light bridge. Its location is indicated with the letter $A$ in Fig. \ref{fig:mapa_GREGOR}. Figure \ref{fig:inversion_umbra} corresponds to the umbral atmosphere marked as $B$ in Fig. \ref{fig:mapa_GREGOR}. In both locations the SIR results show a satisfactory performance, providing an inferred atmosphere that captures the most relevant features present in the observed profiles. The continuum next to the \SiI\ 10827 \AA\ line is slightly below that from the observed profiles. Deep layer temperatures depending on this line might be a bit low. At the umbral position (Fig. \ref{fig:inversion_umbra}), the intensity absorption profile of the \CaI\ 10839 \AA\ line is decomposed into two $\sigma$-components with circular polarization, as can be seen in the Stokes $V$ profile. This fact points to a strong magnetic field in the direction of the LOS. According to the sign of the lobes in Stokes $V$, the polarity of the umbra is negative. On the other hand, at location A the magnetic field is not strong enough to produce a visible Zeeman splitting in the intensity of the \CaI\ 10839 \AA\ line. The reduction of the LOS magnetic field is also evident in the reduced wavelength shift of the Stokes $V$ lobes. Stokes $Q$ and $U$ show significant values, indicating the presence of a considerable horizontal magnetic field.

The determination of the magnetic field perpendicular to the LOS from the linear polarization signals (Stokes $Q$ and $U$) leads to two compatible solutions which differ by 180$^\circ$ in its azimuthal direction. Several techniques have been developed to resolve this ambiguity \citep[\eg,][]{Lites+etal1995, Leka+etal2009}. We have addressed this issue by assuming an approximately radial structure inside the sunspot and carefully chosen the azimuth origin around the center of the spot. The azimuth value retrieved from the inversion was changed by 180$^\circ$ in those regions where it was needed in order to satisfy this assumption.

\section{Results}
\label{sect:results}

The inversion procedures were applied to every spatial position within the FOV of GRIS observations. A median filter was applied to the inversion results to remove the spurious values of the atmospheric parameters retrieved at those individual pixels where the convergence of the inversion is inadequate (at the locations where the differences between the observed and synthetic Stokes spectra resulting from the inversion, quantified by the $\chi^2$, are too high). The value of each parameter at a certain optical depth and spatial coordinates $X$ and $Y$ was substituted by the median in a 3$\times$3-pixel neighborhood.

\subsection{Examination of the atmosphere}
\label{sect:examination}

\begin{figure}[!ht] 
 \centering
 \includegraphics[width=9cm]{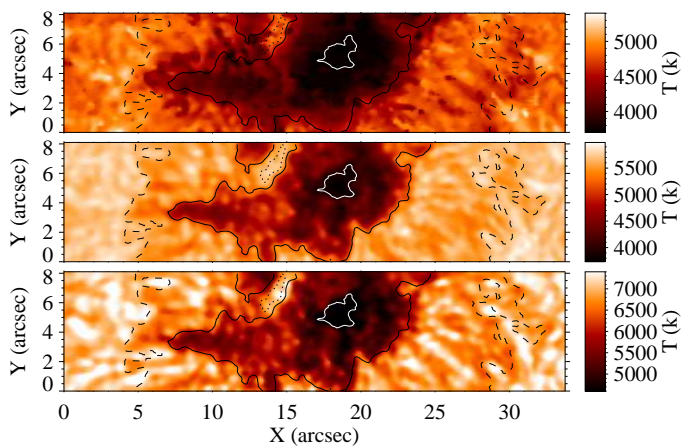}
  \caption{Temperature map inferred from simultaneous inversions of \CaI\ 10839 \AA\ and \SiI\ 10827 \AA\ lines at $\log\tau=-2.2$ (top panel), $\log\tau=-0.5$ (middle panel), and $\log\tau=0.3$ (bottom panel). Contour lines correspond to the same masks as defined in Fig. \ref{fig:mapa_GREGOR}.}
  \label{fig:mapa_t}
\end{figure}

\begin{figure}[!ht] 
 \centering
 \includegraphics[width=9cm]{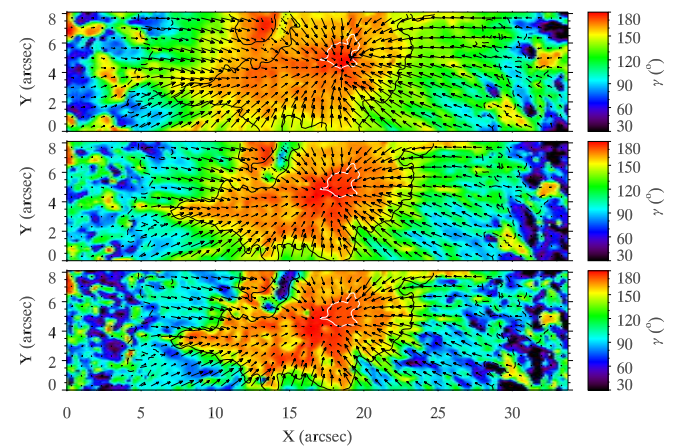}
  \caption{Same as Fig. \ref{fig:mapa_t}  but for magnetic field inclination in the local solar reference frame. Black arrows indicate the direction of the horizontal magnetic field.}
  \label{fig:mapa_Bincl}
\end{figure}

Figures \ref{fig:mapa_t}\,-\,\ref{fig:mapa_Bincl} show maps of temperature, LOS velocity, and magnetic field strength and orientation for the same region and time illustrated in Fig. \ref{fig:mapa_GREGOR}. The contour lines indicate the masks defined in Fig. \ref{fig:mapa_GREGOR}. Maps at three different atmospheric levels, corresponding to $\log\tau=0.3$, $-0.5$, and $-2.2$ are plotted. As pointed out in Sect. \ref{sect:inversion}, the two higher layers correspond to the peaks in sensitivity of the \CaI\ and \SiI\ lines to the magnetic field, while $\log\tau=0.3$ is the deepest layer where we consider that our inversion results are trustworthy.

\begin{figure}[!ht] 
 \centering
 \includegraphics[width=9cm]{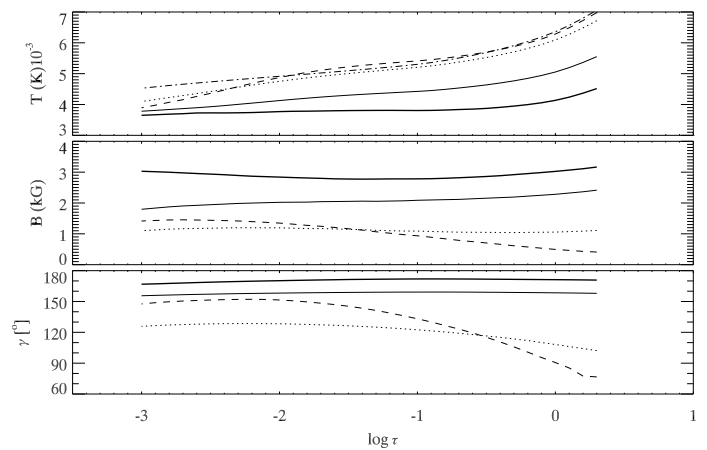}
  \caption{Stratification of the atmospheric parameters averaged over the region delimited by the masks defined in Figure \ref{fig:mapa_GREGOR}. From top to bottom: temperature, magnetic field strength, and LOS magnetic field inclination. Thick solid line: dark umbra; thin solid line: faint umbra, dotted line: penumbra; dashed line: light bridge; dashed-dotted line: quiet Sun (only in top panel).}
  \label{fig:average_atms}
\end{figure}

The general structure of the sunspot is consistent with a vast number of earlier studies \citep[\eg][]{Lites+etal1993b,delToroIniesta+etal1994, Stanchfield+etal1997, WestendorpPlaza+etal2001, BelloGonzalez+etal2005, Langhans+etal2005, Balthasar+Collados2005, BellotRubio+etal2007, Beck2008, Balthasar+Gomory2008, Tiwari+etal2015}. The top panel of Fig. \ref{fig:average_atms} shows the variation with optical depth of the temperature averaged over the regions delimited by the masks defined in Fig. \ref{fig:mapa_GREGOR}. At lower atmospheric heights there is a stark contrast between the temperature of dark umbra, faint umbra, and the other regions. Below $\log\tau=-0.6$ the light bridge shows quiet Sun temperatures and for $\log\tau$ between $-0.6$ and $-2$ its temperature is even higher than that of quiet Sun. At the higher atmosphere the light bridge temperature exhibits a steep gradient, reaching an umbral value around $\log\tau=-3$.

\begin{figure}[!ht] 
 \centering
 \includegraphics[width=9cm]{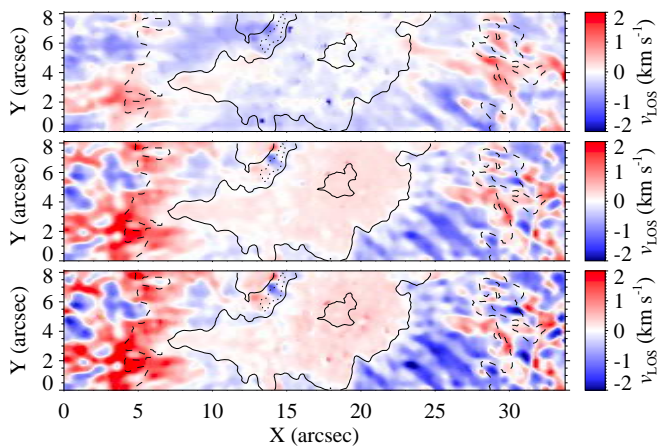}
  \caption{Same as Fig. \ref{fig:mapa_t} but for LOS velocity. Color scale is saturated between -2 and 2 km s$^{-1}$.}
  \label{fig:mapa_vLOS}
\end{figure}

The two deeper velocity maps from Fig. \ref{fig:mapa_vLOS} show clearly the Evershed flow \citep{Evershed1909}. Interestingly, at $\log\tau=0.3$ the light bridge shows an upflow with a velocity of $-1$ km s$^{-1}$ surrounded by two downflows, the strongest with a velocity of 2 km s$^{-1}$. These flows are present only in the side of the light bridge next to the small umbra. The downflows decrease with height and completely vanish by $\log\tau=-2.2$ (top panel). The upflow keeps its value of around $-1$ km s$^{-1}$ for all the optical depths analyzed.

\begin{figure}[!ht] 
 \centering
 \includegraphics[width=9cm]{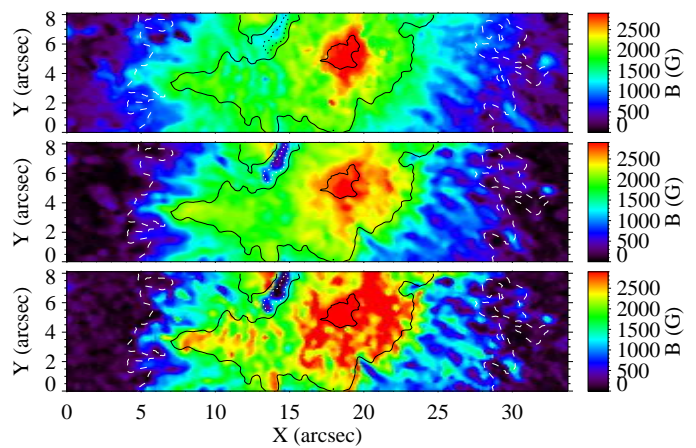}
  \caption{Same as Fig. \ref{fig:mapa_t} but for magnetic field strength.}
  \label{fig:mapa_Bstrength}
\end{figure}

The atmospheres obtained from the inversion are referred to a system of reference with the vertical directed along the LOS. Due to the position of the active region on the solar disk, this reference frame does not coincide with the local reference frame, where vertical magnetic fields are radially oriented. The magnetic field has been transformed to the solar local reference frame. The strength and orientation of the magnetic field are illustrated in Figs. \ref{fig:mapa_Bstrength} and \ref{fig:mapa_Bincl}. At the three plotted optical depths, the photospheric magnetic field ranges from 3000 G strength and a nearly vertical orientation at the center of the umbra to 1000 G and more horizontal inclination at the penumbra. At the deepest layer ($\log\tau=0.3$) the inclination of the penumbral magnetic field is higher, showing values around 90$^{\circ}$ and even reversals.

At the lower atmosphere the light bridge shows a strong reduction of magnetic field strength, with the lowerst trustworthy value as low as 30 G, and becomes more horizontal. At some locations this low magnetic field results in a polarity reversal. The light bridge average field inclination at the deepest layer is 76$^\circ$, but at certain locations its inclination is around 25$^\circ$. Figure \ref{fig:average_atms} shows that for heights below $\log\tau=-0.5$ the magnetic field in the light bridge is more horizontal and has lower strength than the penumbral magnetic field. At higher levels its strength increases to above that of the penumbral average and at the same time it becomes more vertical. The configuration of the magnetic field in the light bridge around $\log\tau=-2.2$ is closer to the faint umbra than the penumbra, according to its inclination of around 150$^\circ$ and its higher strength. The topology of the magnetic field in the light bridge can be pictured as a region with low horizontal magnetic field (including strong reversals in the magnetic field lines) covered by a canopy of stronger and more vertical field, similar to the surrounding umbral field \citep{Jurcak+etal2006, Lagg+etal2014,Schlichenmaier+etal2016}.

\subsection{Geometrical scale}

The optical depth scale ($\log\tau$) retrieved with SIR has been converted to a geometrical scale ($z$) assuming hydrostatic equilibrium. At each location, the hydrostatic equilibrium has been reevaluated taking into account magnetic pressure, but neglecting magnetic tension. Next, the geometrical height has been obtained from the new density and the opacity evaluated at a continuum wavelength. Due to the Wilson depression, the reference height of the geometrical scale for each position is different. We have chosen to define $z=0$ at the height of $\log\tau=0$ in the quiet Sun atmosphere. All the inverted atmospheres have been shifted in the vertical direction in order to look for the configuration that satisfies the pressure balance. In the sunspot this is given by \citep[\eg,][]{Mathew+etal2004}:

\begin{equation}
P_0(z)=P_\mathrm{g}(r,z)+B_\mathrm{z}^2(r,z)/(2\mu_0),
\label{eq:MHS}
\end{equation}

\noindent where $P_0(z)$ is the gas pressure stratification in the quiet Sun, $P_g(r,z)$ is the gas pressure in the spot, $B_z(r,z)$ is the vertical magnetic field, and $\mu_0$ is the magnetic permeability. The curvature force has been neglected, similarly to previous works \citep[\eg][]{SanchezAlmeida2005, Mathew+etal2004}. For a discussion of the influence of magnetic tension on the estimation of the Wilson depression see \citet{MartinezPillet+Vazquez1993} and \citet{Solanki+etal1993}. We have determined $P_0(z)$ as the average of the gas pressure over all the quiet Sun surrounding the sunspot. The Wilson depression $\Delta z_\mathrm{WD}(x,y)$ is determined as the shift in $z$ which needs to be applied to the total pressure (gas and magnetic pressure) at every position $[x,y]$ in order to match $P_0$ at $\log\tau=0$ (that we have set as $z=0$ in the quiet Sun). This method proved to be problematic when determining $\Delta z_\mathrm{WD}(x,y)$ in the umbra and regions with high magnetic field strength. In those cases, for all the heights where the response functions of the spectral lines are large enough, the total pressure in the magnetized atmosphere was higher than $P_0(z=0)$. This can be partially due to the fact that the gas pressure retrieved from the inversion might be inaccurate in the magnetized regions, since the inversion code does not include the magnetic pressure term in the derivation of the pressure scale. We have determined the change in the Wilson depression of every position with respect to an arbitrarily chosen location with a moderate field strength, instead of the average quiet Sun. In the quiet Sun atmosphere, the range of heights where the inversion provides a reliable result spans around 450 km, while typical Wilson depressions can be higher than 600 km. This means that in some umbral regions the vertical displacement required for matching the photospheric pressures is higher than the available atmosphere. The total pressure stratification of the reference atmosphere is between the low (quiet Sun) and strong (umbra) field cases and the geometrical shift can be directly obtained at all locations. The absolute Wilson depression has been then retrieved by adding the shift between the average quiet Sun and the reference position. 

Our estimation of the Wilson depression has some limitations. First, we have neglected the magnetic tension. A rigorous determination would require solving the magnetohydrodynamic equations on a three dimensional grid. The motion equation could be solved even including the velocity and its gradients, although in this case we would miss information about the horizontal components and the temporal derivatives. In addition, the stratification of the parameters retrieved from the inversion imposes some restrictions that will prevent us from obtaining a physical solution due to the uncertainties in the determination of the magnetic field out of the optimal sensitivity region of the spectral lines. Accounting for all this complexity is a huge challenge and is out of the scope of this paper \citep[see][for an example where the Lorentz force is included in the force balance equation]{Puschmann+etal2010}. Finally, we have obtained $\Delta z_\mathrm{WD}(x,y)$ by imposing pressure balance at one specific height. The balance is not satisfied at all $z$ and, thus, choosing a different height as a reference would lead to some differences in the Wilson depression.

\begin{figure}[!ht] 
 \centering
 \includegraphics[width=9cm]{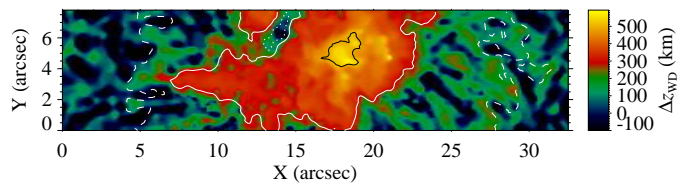}
  \caption{Wilson depression of the observed sunspot. Contour lines correspond to the masks defined in Figure \ref{fig:mapa_GREGOR}.}
  \label{fig:WD}
\end{figure}

Figure \ref{fig:WD} illustrates the Wilson depression inferred for the observed sunspot. The maximum value of the Wilson depression is almost 600 km. We obtain average values for $\Delta z_\mathrm{WD}(x,y)$ around 500 km for the dark umbra, 330 km for the faint umbra, 110 km for the penumbra, and 8 km for the light bridge. The light-bridge photosphere is basically at the height of the quiet Sun photosphere and at some point its Wilson depression is -100 km, that is, the surface at those locations is above the average surface of the quiet Sun. The contrast in $\Delta z_\mathrm{WD}(x,y)$ of the light bridge with respect to the dark surrounding regions is in the range 200-400 km, in agreement with previous measurements \citep[\eg,][]{Lites+etal2004}.

\subsection{Three-dimensional structure}
\label{sect:3D}
\begin{figure*}[!ht] 
 \centering
 \includegraphics[width=14cm]{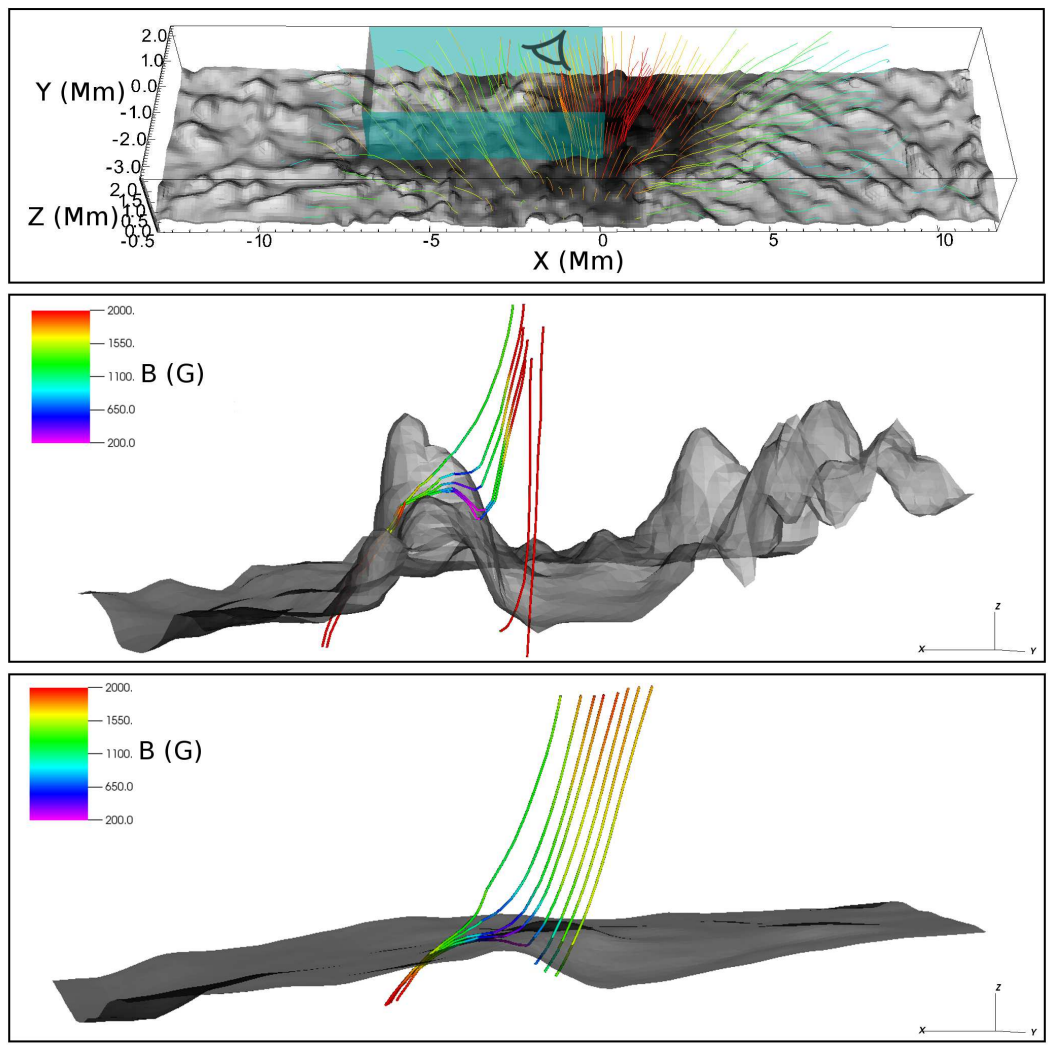}
  \caption{Three-dimensional structure of the sunspot in the local reference frame. Continuum intensity is plotted at constant $\log\tau=0$ in grey scale. Magnetic field lines are illustrated with a color scale showing the magnetic field strength at the corresponding location from lower than 200 G (purple) to higher than 2000 G (red). Middle panel: detailed view of the light bridge region (delimited by the light blue box in the top panel and from the point of view indicate by the eye in top panel) and some selected field lines obtained from the analysis of the deconvolved maps (see Section \ref{sect:PSF}); bottom panel: detailed view of the light bridge region (same region illustrated in middle panel) and some selected field lines obtained from the analysis of the original non-deconvolved maps (see Sect. \ref{sect:PSF}). In the middle and bottom panels the $z$-axis has been expanded by a factor three.}
  \label{fig:3D}
\end{figure*}

The two-dimensional maps of the solar surface scanned during the observations and the depth-dependent stratification of the inferred physical parameters obtained with SIR allow us to construct a three-dimensional model of the observed active region. Figure \ref{fig:3D} illustrates the thermodynamic and magnetic structure of the sunspot in the local reference frame after transforming the optical depth scale of the stratification into a geometrical scale taking into account the Wilson depression. All the parameters are plotted for $\log\tau$ lower than 0.3, so that we are only including the region where the sensitivity of the spectral lines is high enough.

The top panel of Fig. \ref{fig:3D} shows the region presented in Figs. \ref{fig:mapa_t}\,--\,\ref{fig:mapa_Bincl}, with the horizontal direction expressed in megameters and the position $(x=0,y=0)$ corresponding to the center of the umbra. The continuum intensity is plotted at the surface $\log\tau=0$. Magnetic field lines exhibit a radial geometry. Their color scale illustrates the magnetic field strength at the corresponding location. At the dark umbra magnetic field lines are mostly vertical and their strength is above 2000 G (red). As farther regions from the center of the sunspot are considered, the magnetic field strength is reduced and its inclination is increased until it reaches the penumbra, where comparatively horizontal magnetic field lines dominate. In this panel, the field lines are plotted only where the magnetic field strength is above 500 G.

For a more detailed analysis of the light bridge, the middle panel of Fig. \ref{fig:3D} shows an expanded plot of the domain delimited by the light blue box in the top panel. The point of view of the middle picture is indicated by the ``eye'' in the top panel. In this plot, the $z$-scale has been increased by a factor of 3 in order to better illustrate the Wilson depression. Some selected field lines that cross the light bridge have been plotted in order to illustrated the magnetic field configuration. The center of the umbra is pushed down almost 600 km with respect to the quiet Sun surface. The light bridge shows a remarkable variation of the height at which the continuum intensity is formed, since it comes from a layer between 200 and 400 km higher than the $\log\tau=0$ height at both sides. A movie showing the three-dimensional structure of the light bridge is available in the online material.

\begin{figure}[!ht] 
 \centering
 \includegraphics[width=9cm]{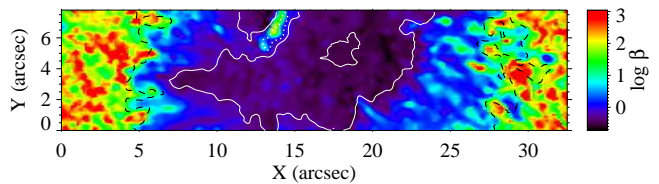}
  \caption{Plasma-$\beta$ at $\log\tau=0$ in logarithmic scale. Contour lines correspond to the masks defined in Figure \ref{fig:mapa_GREGOR}.}
  \label{fig:beta}
\end{figure}

Figure \ref{fig:beta} illustrates the plasma-$\beta$ at a surface with constant optical depth $\log\tau=0$ on a logarithmic scale. In the umbra (dark and faint), the isosurface $\log\tau=0$ lies above the $\beta=1$ surface, indicating that the continuum is formed in a low-$\beta$ regime, where magnetic pressure is higher than gas pressure. The photospheric umbra is dominated by magnetic field with plasma-$\beta$ values as low as 0.2. At the penumbra the photosphere change from a magnetic dominated regime at the umbra-penumbra boundary to a plasma dominated regime at the penumbra-quiet Sun boundary. Strong variations in the plasma-$\beta$ associated with penumbral filaments are found at the west side of the penumbra. On the one hand, bright penumbral filaments exhibit plasma-$\beta$ roughly between 10 and 1000. On the other hand, plasma-$\beta$ in dark penumbral regions is around unity. The light bridge shows a striking increase in the plasma-$\beta$. The small umbra at the east side of the light bridge is dominated by magnetic pressure, with a plasma-$\beta$ around 0.2. However, in the light bridge the plasma-$\beta$ is in the range 1-200.

The dynamics of the light bridge are clearly dominated by plasma motions and this fact has a critical relevance for reshaping the magnetic field. The aforementioned velocity flows, consisting of a 1 km s$^{-1}$ upflow surrounded by two downflows with a magnitude of 2 km s$^{-1}$ at $\log\tau=0.3$ and decreasing with height, indicates that the light bridge exhibits vigorous convection. Since in the light bridge the magnetic field is frozen to the plasma motion, magnetic field lines are bent and become more horizontal just on top of the $\log\tau=0$ surface. At some locations they are even completely reversed, exhibiting a change in the polarity of the magnetic field. The bending of the field lines produced by the convective motions is cleary visible at the purple section of the field lines plotted in the middle panel of Fig. \ref{fig:3D}. 

The small umbra at the east side of the light bridge imposes a strong almost vertical magnetic field (red vertical field lines in the middle panel of Fig. \ref{fig:3D}). When the field lines crossing the light bridge reach those originating at the small umbra, the magnetic field tension is strong enough to reverse again the direction of the field lines. At those locations, they point in the same direction as they did before crossing the light bridge. Field lines return to a close to vertical configuration, joining at some height the magnetic field lines from the other side of the light bridge (middle panel of Fig. \ref{fig:3D}). 

Across the light bridge the magnetic field strength is significantly reduced. The variation of the magnetic field (in comparison with the surrounding dark umbra) decreases with height and goes from around $\Delta B_\mathrm{ch}=2200$ G at $z=0$ to $\Delta B_\mathrm{ch}=1000$ G at $z=1$ Mm. As pointed out before, magnetic field lines form a canopy. Below this canopy the magnetic field strength is even lower.

\subsection{Original vs deconvolved maps}
\label{sect:original_deconvolved}

The analysis of the sunspot has been repeated using the original maps instead of those obtained from the deconvolution discussed in Sect. \ref{sect:PSF}. The deconvolution allowed us to take into account the PSF of the telescope (including the effect of seeing) and correct the stray light contamination. 

In both cases we obtain a similar global structure. The umbral and penumbral magnetic field show a radial structure. Regarding the light bridge, both cases point to a cusp-like magnetic field configuration, with a lower magnetic field below the canopy and a more vertical field at higher layers. However, there are also significant differences. The canopy obtained from the original maps is formed by mainly horizontal magnetic fields. Some reversals in the field polarization are also measured and small deviations of the field lines from the horizontal are found. However, their bending is barely visible in the bottom panel of Fig. \ref{fig:3D}, which shows results obtained from the analysis of the original data for the same region plotted in the middle panel of that figure. On the contrary, the deconvolved map shows clear signatures of the twisting of the field lines produced by convective motions.

These light bridge convective motions, clearly visible in the velocity map retrieved from the analysis of the deconvolved map, are absent in the flow inferred from the original map. The later only exhibits a 2 km s$^{-1}$ downflow at the side of the light bridge next to the small umbra.

Our estimation of the Wilson depression in the deconvolved map (around 600 km in the umbra) is higher than that obtained by \citet{Mathew+etal2004} using a similar approach (around 400 km) and agrees with that retrieved from geometric techniques using the Wilson effect \citep[600$\pm$200 km][]{Gokhale+Zwaan1972}. \citet{Mathew+etal2004} consider that the difference in their derived Wilson depression is due to neglecting the magnetic tension. However, from the analysis of the original map, the maximum Wilson depression that we obtain in the umbra is slightly higher than 300 km. This result is closer to the value given by \citet{Mathew+etal2004}, and significantly lower that the one inferred with a better resolved map.

The magnitude of the Wilson depression inferred from the original data is strikingly lower than that from the deconvolved map. This effect is specially evident at the light bridge. The height of the $\log\tau =0$ surface around the light bridge in the deconvolved and original maps is illustrated in the middle and bottom panels of Fig. \ref{fig:3D}, respectively. The height difference between the low field strength regions and their surroundings retrieved from the analysis of the original map is below 100 km, while in the deconvolved map this jump is up to 400 km, in agreement with \citet{Lites+etal2004}.

\section{Discussion and conclusions}
\label{sect:conclusions}

We have analyzed the magnetic topology and thermodynamic structure of parts of a sunspot in active region NOAA 12096 a few days before it completely decayed. Our analysis has been focused on the photosphere, with special attention to the magnetic configuration of the deep photosphere that can be explored thanks to the \CaI\ 10839 \AA\ line. This spectral line is sensitive to deeper layers than other commonly used photospheric lines. 

The inversion of the active region reveals many well known and widely explored sunspot properties. Among other features, we have shown the temperature structure and stratification of the sunspot atmosphere, traces of the Evershed flow, changes in the magnetic field from strong and vertical in the umbra to weaker and mostly horizontal in the penumbra, and Wilson depression all over the active region. The most compelling feature for further exploration is the deep photosphere in the light bridge that was formed between the main umbra and a nearby small umbral region. The magnetic field at this location is not only much lower than in the surrounding umbral regions, but also lower than the penumbral magnetic field. At the same time, its inclination with respect to the local vertical is also lower, showing horizontal field lines and even reversals in their trajectory. The vertical stratification of the light bridge atmosphere is clearly different from other penumbral regions.

The magnetic field structure of the light bridge reveals several similarities with the configuration that has been found for previously analyzed light bridges: it exhibits lower and more horizontal magnetic field relative to the umbra \citep{Beckers+Schroter1969, Lites+etal1991,Leka1997}, its brightness is similar to the penumbra and it separates two regions with the same polarity \citep{Sobotka+etal1993, Rimmele2008}, the higher layers are comparable to the surrounding umbral atmosphere \citep{Ruedi+etal1995}, and the field lines form a cusp-like structure with a reduced field region in the inner part \citep{Leka1997, Jurcak+etal2006, Lagg+etal2014}. In addition, the velocity profile shows central hot upflows surrounded by cooler downflows. This flow pattern has been detected in other light bridges and it is interpreted as a evidence for vigorous convection in the light bridge \citep{RouppevanderVoort+etal2010, Lagg+etal2014}.

The flow pattern measured for the light bridge exhibits some differences with respect to those observed in previous analyses. In this light bridge we do not detect a complete convective cell, where the upflow is completely surrounded by downflows. Instead, we have only detected downflows at the side of the light bridge closer to the small umbra. The width of this cell (1.9\arcsec) is comparable to the typical width of quiet-Sun granules and those observed in light bridges \citep[$1-2$\arcsec][]{Lagg+etal2014}. Only one of these convective cells is found in the analysed light bridge, whose convective motions resemble those from quiet-Sun granules, except for the asymmetry in its downflows. The light bridge is relatively small, and it is possible that missing convective downflows are spatially unresolved near the umbral edges of both sides of the light bridge. Pressure and horizontal motions towards the edges of the light bridge could squeeze these downflows to very small spatial scales, out of the reach of the spatial resolution achieved even with the deconvolution described in Sect. \ref{sect:PSF}, since the small-scale stray light could not be properly characterized. We also find some signatures of convection in the flow pattern at the centre side of the penumbra (see the range x=[25\arcsec, 28\arcsec] and y=[6\arcsec, 8\arcsec] from bottom panel of Fig. \ref{fig:mapa_vLOS}). The presence of convective motions in the penumbra has been pointed out by previous observational works \citep{Ichimoto+etal2007,Zakharov+etal2008, Joshi+etal2011, Scharmer+etal2011, Tiwari+etal2013}. 

As discussed in Sect. \ref{sect:evolution}, the light bridge is created by the motion of a small umbral region towards the main umbra (see movie). It is composed of the penumbral and quiet Sun atmosphere that remained confined between two highly magnetized regions at the time the analysed data were acquired. As the two magnetized regions approach each other, the field lines that spread from the central umbra are blocked by those coming from the dark pore-like feature. The field lines above the light bridge are forced to become more vertical than in the rest of the penumbra (see bottom panel of Fig. \ref{fig:average_atms} around $\log\tau=-2$). This process forms a magnetic canopy. An atmosphere with very low magnetic field and quiet Sun temperature (see top and middle panel of Fig. \ref{fig:average_atms}) remains covered by a cusp-like field configuration. Convection keeps working, but the flows are confined by the low-$\beta$ atmosphere of the surrounding umbrae. Below the canopy, gas pressure is much higher than magnetic pressure (plasma-$\beta$ in the range 10-200). The downflowing material due to convection can drag down towards the interior the magnetic field below the canopy, bending the field lines and leading to reversals in the field orientation. Closer to the small umbra the magnetic tension can reverse again the field lines, which recover an almost vertical configuration and join the vertical field lines from the other side of the umbra.The magnetic canopy confines the plasma to the lower atmosphere. An example of this phenomena was recently found by \citet{Louis+etal2015}, who reported the emergence of a small loop in the lower atmosphere of a sunspot light bridge. The loop does not reach the chromosphere since the magnetic field configuration impedes it from continuing rising.

Field lines reversals in light bridges were recently detected by \citet{Lagg+etal2014}. They proposed two possible scenarios: the field lines reverse their direction again due to magnetic tension, or the field is dragged towards the interior and the field lines are then downward directed. Our results have confirmed the presence of bending in the field lines and have shown evidence that their first scenario is the most plausible. The detection of the two successive reversals in the field lines is only possible thanks to the high sensitivity of the \CaI\ 10839 \AA\ line to the deep layers and the high spatial resolution of GREGOR telescope. Many works have reported chromospheric activity phenomena above light bridges \citep[\eg,][]{Roy1973,Asai+etal2001,Berger+Berdyugina2003,Shimizu+etal2009,Toriumi+etal2015,Robustini+etal2016}. Our observations in \HeI\ 10830 \AA\ do not show any trace of these dynamics events. However, the magnetic configuration found in our work at the deep photosphere can trigger those phenomenon through reconnection of the reversed field lines.

The inversions carried out in this paper have been independently performed using the original map and the spatially deconvolved map, which takes into account the telescope and seeing PSF and the stray light contamination. Both results show a similar general picture. However, the inversion of the original map also present significant differences with the analysis of the deconvolved map. First, it does not reveal a clear pattern of the overturning convective motions of the light bridge. Instead, only a downflow is obtained. Second, the complex bending of the field lines in the light bridge is not undoubtedly detected. Opposite polarity fields are also found in the original non-deconvolved maps, but the curvature of the field lines is much lower. Finally, the magnitude of the Wilson depression is greatly underestimated. Our results indicate the remarkable relevance of correcting for telescope resolution and stray light in order to infer the magnetic and thermodynamic structure of the Sun from spectropolarimetric observations.

The decay of a sunspot is a long process during which the structure of the atmosphere shows a striking and complex evolution. Our analysis revealed the magnetic field configuration of the lower photosphere at a specific time close to the final stages of a sunspot. A comprehensive study of the magnetic configuration during the decay will require tracking an active region over several days. This study would also benefit from the inclusion of many active regions, sampling each stage of the evolution at different positions on the solar disk in order to analyze the magnetic and thermodynamic configuration under different viewing geometries.

\begin{acknowledgements} 
The authors thank Basilio Ruiz Cobo and Carlos Quintero Noda for advice on inverting the data and providing very useful suggestions. The 1.5-meter GREGOR solar telescope was built by a German consortium under the
leadership of the Kiepenheuer-Institut f\"ur Sonnenphysik in Freiburg with the
Leibniz-Institut f\"ur Astrophysik Potsdam, the Institut f\"ur Astrophysik
G\"ottingen, and the Max-Planck-Institut f\"ur Sonnensystemforschung in G\"ottingen as
partners, and with contributions by the Instituto de Astrof\'isica de Canarias and
the Astronomical Institute of the Academy of Sciences of the Czech Republic. SDO HMI data are provided by the Joint Science Operations Center - Science Data Processing. The support of the European Commission's FP7 Capacities Programme under Grant Agreement number 312495 ``SOLARNET'' is acknowledged. This research has been funded by the Spanish MINECO through grant AYA2014-55078-P.
\end{acknowledgements}

\bibliographystyle{aa} 
\bibliography{biblio.bib}

\end{document}